\documentstyle
[12pt]
{ioplppt}
\begin{document}
\jl{3}
\letter{Hall Effect in the mixed state of moderately clean superconductors}
\author{I Vekhter and A Houghton}
\address{Department of Physics, Brown University, Providence, RI 02912-1843,
USA}

\begin{abstract}
   The Hall conductivity in the mixed
state of a clean ($l \gg \xi_0$) type-II s-wave superconductor
is determined from a microscopic calculation within a
quasiclassical approximation. 
We find that
 below the superconducting transition  the contribution 
to the transverse conductivity due to dynamical 
fluctuations of the order parameter
is compensated by the modification of the quasiparticle contribution.
In this regime the nonlinear behaviour of the Hall angle is governed by the 
change in
the effective quasiparticle scattering rate 
due to the reduction in the density of states at the Fermi 
level.
The connection with experimental results is discussed.

\end{abstract}

\pacs{74.25.Fy, 74.60.-w, 74.60.Ge}

\maketitle

The Hall effect in the mixed state of type-II superconductors
has remained a theoretical puzzle 
for almost thirty years \cite{Dorsey}.
The existing phenomenological \cite{BS,NV}  
theories predict that the Hall angle in the flux-flow regime is
either identical to that in the normal state \cite{BS} or constant \cite{NV}.
Theories which make use of the time-dependent Ginzburg-Landau equations (TDGL)
also find no modification of the
Hall conductivity in the superconducting state \cite{Dorsey}.
These predictions are at variance with the
strongly nonlinear behaviour (as a function of magnetic field) 
found in experiments performed on both low-$T_c$ materials
\cite{Hagen1,Fiory} and high-$T_c$ cuprates \cite{Hagen1,Ong}.
For dirty superconductors ($l \ll \xi_0$, where $l$ is the mean free path and
$\xi_0$ is the superconducting coherence length),
transport coefficients can be determined from microscopic
theory by a straightforward expansion in powers of the order
parameter, $\Delta$.
The results of such a calculation for the transverse resistivity
\cite{Ebis}
explain qualitatively the sharp increase in the Hall angle observed in 
experiment (although, to our knowledge, no systematic comparison has
been made), 
and provide the physical basis for a generalized TDGL approach, in which
the relaxation rate is assumed to be complex, rather than purely real,
to allow for a modification of the transverse transport coefficients
\cite{Dorsey,makihall}. 
The small parameter in the expansion of the
microscopic equations is
proportional to both the order parameter and the mean free path, 
therefore, it is not small 
in the clean  ($l \gg \xi_0$) 
limit. In this regime a straightforward expansion is not possible
and the TDGL equations are not applicable \cite{HM}, so that an
alternative approach is needed for the calculation of the transverse
transport coefficients.

In this Letter we give the results of 
a calculation of the Hall conductivity
of a clean s-state superconductor in the mixed state
near the upper critical field, $H_{c2}$,
which uses a quasiclassical approximation to the microscopic theory.
 We made this choice
as both the normal state and superconducting properties of the
low-$T_c$ compounds are well known, and comparison between 
theory and experiment is fraught with less ambiguity.
The quasiclassical approach
\cite{E} has been applied successfully in the past
to study transport
phenomena in superfluids \cite{Rainer1} and superconductors \cite{Ov}
and more recently to the unconventional superconductors
\cite{Sauls}. 
  The central quantity in this method is the single-particle matrix 
Green's function $\widehat G$
integrated over the quasiparticle energy
\begin{equation}
\widehat g(s,{\bf R}; \omega_n,\omega_{n^{\prime}})= 
                                       \pmatrix{g& -f\cr
                                         f^{\dag} & \bar{g}\cr}=
    \int { d \zeta_p \over \pi}
                \widehat 
                G(p,{\bf R}; \omega_n,\omega_{n^{\prime}});
\label{defg}
\end{equation}
here the
 $\omega_n= 2 \pi T (n+{1 \over 2})$ are fermionic Matsubara
frequencies, 
$s$ is the normalized parameterization of the Fermi surface,
${\bf R}$ is the
center of mass coordinate
 and $p$ is the relative momentum.
Since  the Green's function is strongly peaked at the Fermi momentum $p_f$, 
which is
normally far larger than any other momenta in the problem,
slower varying quantities 
such as the self-energy and external potential can be expanded around their
values at the Fermi surface. The result of such an expansion \cite{Rainer1}
in the small parameter
$1/{\xi_0 k_f} \sim \Delta / \epsilon_f$ 
is a set of transport-like equations for the
quasiclassical propagator $\widehat g$.
   We have generalized these equations to include terms responsible for
the Hall Effect in a charged superfluid. Technical details of the
derivation will be reported elsewhere \cite{us}, here we use the equations
to determine the 
transverse dc-conductivity

We use linear response theory in the 
vector
potential ${\bf {A}}(\omega_0)$ describing a constant electric field
${\bf {E}}=E\widehat{\bf x}$.
The magnetic field  
${\bf {H}}=H \widehat {\bf z}$, chosen parallel to the z-axis,
 is described by the vector potential
${\cal A}({\bf {R}})=Hx\widehat {\bf y}$.
 We consider
a spherical Fermi surface and use the Born approximation for
 s-wave impurity scattering characterized by
 a collision rate $\tau^{-1}$.
The spatial dependence of the order
parameter is modeled by the periodic Abrikosov vortex lattice 
\begin{equation}
\label{VL}
\Delta({\bf R})=\sum_{k_y} C_{k_y} e^{i k_y y} \Phi_0 (x - \Lambda^2 k_y).
\end{equation}
Here  $\Phi_0 (x)$ is the lowest energy eigenfunction of the linearized
Ginzburg-Landau equation (i.e. the eigenfunction of a harmonic oscillator
with Cooper pair mass $M=2m$ and frequency $\omega_c$) and
 $\Lambda^2=(2eH)^{-1}$ is the magnetic length ( $\Lambda \sim \xi_0$
for fields $H \sim H_{c2}$).
  This approach is appropriate provided that the broadening of the levels 
in the
vortex core is large compared to 
their spacing 
$1/ \tau \gg \Delta^2 /
\epsilon_f$, it breaks down in the superclean regime (cf. \cite{kop}).
In the clean limit the finite lifetime is accounted for by replacing 
$\omega_n$ by 
$\widetilde \omega_n= \omega_n+
(2 \tau)^{-1}\langle g(\widetilde \omega_n)\rangle$ 
(angular brackets denote an average over the Fermi surface);
 corrections to the order parameter due to
impurity renormalization are of the order
$O(\Lambda / l)$ and can be ignored.
The  equations for the unperturbed functions $f$ and $g$
and the linear, in ${\bf A}$, corrections to the propagator $f_1$ and $g_1$ are
\cite{us}
\begin{eqnarray}
\label{f0}
\fl
\bigl[2 \widetilde \omega_n  + {\bf v}_f 
( \nabla - 2 i e {\cal A})\bigr] f =
2 i \Delta g
\\
\fl
\label{f1}
\bigl[ 2 \widetilde \Omega_n + {\bf v}_f (\nabla - 2 i e {\cal A})\bigr] f_1 
          =
            i e {\bf v}_f {\bf {A}} (f + f(-))
                 + i \Delta (g_1 - \bar g_1) +
                          i \Delta_1 (g + g(-))
\\
\fl
\label{g1}
\bigl(i \widetilde \omega_0 + i \omega_c{\partial \over \partial \phi}\bigr) 
\bigl(g_1- \bar g_1\bigr)=
              2 e {\bf v}_f {\bf A} (g - g(-)) + 
              \Delta_1^{\star} \bigl(f-f(-)\bigr)+ 
              \Delta_1 \bigl(f^{\dagger}- f^{\dagger}(-)\bigr)
\\
\nonumber \qquad \qquad
+(2\tau)^{-1} \biggl( \langle f_1^{\dagger} \rangle
		\bigl(f-f(-)\bigr) + \langle f_1 \rangle 
                               \bigl(f^{\dagger}- f^{\dagger}(-)\bigr)
					\biggl)
\\
\nonumber \qquad \qquad
-{i \over 2}\biggl[ {\partial \Delta_1^{\star} \over \partial {\bf R}}
		    {\partial \over \partial {\bf p}_{||}}
		   \bigl(f+f(-)\bigr)
                 +{\partial \Delta_1 \over \partial {\bf R}}
		  {\partial \over \partial {\bf p}_{||}}
                   \bigl(f^{\dagger}+f^{\dagger}(-)\bigr) 
\\
\nonumber \qquad \qquad \qquad
	       +2{\partial \Delta^{\star} \over \partial {\bf R}}
		    {\partial f_1^{\dagger} \over \partial {\bf p}_{||}}
		+2{\partial \Delta \over \partial {\bf R}}
		{\partial f_1	 \over \partial {\bf p}_{||}}	
    \biggr],
\end{eqnarray}
\noindent
where $\Delta_1$ is the change in the order parameter induced by the
electric field.
There  are corresponding equations for $f^{\dagger}$ and $\bar g$.
In these equations  $ \sigma_z $ is the Pauli matrix,
the Fermi velocity
${\bf v}_f(s) = v (\sin \theta \cos \phi, 
                        \sin \theta \sin \phi, 
                        \cos \theta)$,
${\bf p}_{||}$ is the component of the momentum parallel to the Fermi surface,
$\omega_c = e H / m c$ is the cyclotron frequency, 
 and  
the angular brackets denote an average over the Fermi surface.
We have used the shorthand notations
${\omega_n-}=\omega_n - \omega_0$ and
$g=g(\widetilde \omega_n)$, 
$g(-)=g(\widetilde {\omega_n-})$, and introduced
$2 \widetilde \Omega_n = \widetilde \omega_n + \widetilde {\omega_n - }$ ,
$\widetilde \omega_0 = \widetilde \omega_n - \widetilde {\omega_n -}$.
The superconducting order parameter
satisfies the usual self-consistency condition
$\Delta({\bf R})= {\rm g} N(0) \pi T \sum_n \int d^2s f(s,{\bf R};
                                                      \omega_n,\omega_n)$,
where g is the coupling constant and $N(0)$ is the density of states at the 
Fermi
surface. 
 The unperturbed function $\widehat g$ obeys the normalization conditions 
$g+\bar g=0$ and 
 $g^2 - f f^{\dagger} = -1$\cite{E}.
We chose to write an equation for $g_1 - \bar g_1$ since the transport current is given by 
\cite{lar}
\begin{equation}
\label{curr}
{\bf j} = {1 \over 2}
             \pi e N(0)T \sum_n \int d^2 s \  {\bf v}_f (s) 
           \bigl(g_1(s,\omega_n)- \bar g_1(s,\omega_n)\bigr).
\end{equation}
\noindent
and the functions $f_1$, $f_1^{\dagger}$ depend on $g_1$, $\bar g_1$
in this combination only.
In equations (\ref{f0})-(\ref{g1})
we have omitted terms whose contribution to the conductivity is
of order $O(\Lambda / l)$ smaller than that of leading order
terms.
Equation (\ref{f0}) is the well-known static Eilenberger equation
\cite{E} and equations (\ref{f0})-(\ref{g1}) contain  all the terms 
relevant to the Hall effect for a clean type-II
superconductor in the high-field region.
It is clear from equation (\ref{g1}) that there are several distinct 
contributions to $g_1 - \bar g_1$ (and
therefore the current). The first, the 
quasiparticle contribution, depends on the unperturbed function $g$,
 while the
second, proportional to $\Delta_1$, 
is due to the dynamical fluctuations of the order parameter induced by the
perturbing electric field. 
The third term in this equation
describes the additional scattering of 
quasiparticles off these dynamic fluctuations; it has the same origin
as the Thompson diagram in the analysis  of transport 
in dirty superconductors \cite{Thom}.  The fourth term describes how
 as vortices move and are
deformed by the transport current, the resulting gradients of the 
order parameter act as  driving forces in the transport-like equations. 
Finally, as the renormalization
of the
frequency $\omega_0$  in equation(\ref{g1}) depends on the angular average of 
the quasiparticle
propagator, which changes in the superconducting state,
the effective transport mean free path is modified. 
  
    To proceed we approximate 
the diagonal part of the quasiclassical propagator by its spatial
average \cite{BPT}.
Since the 
electromagnetic fields in a superconductor vary over 
distances of the order of the
penetration depth $\lambda$, this is a very good approximation in the London
limit $\kappa=\lambda / \xi_0 \gg 1$;
even for compounds with moderate values of  $\kappa$
it remains valid for a wide field range below $H_{c2}$.
In all of the following $g$ will stand for the averaged distribution function.
We now solve equation(\ref{f0}) with the normalization condition to determine
the unperturbed functions $f$ and $g$. We then determine $f_1$ and
 $\Delta_1$ to leading order in $(\Lambda \Delta /v)^2 \ll 1$ by
solving equation (\ref{f1}) together with the self-consistency 
condition for the order 
parameter. 
To accomplish this program 
 the expression               
 $(2 \widetilde \omega_n + {\bf v}_f \cdot 
( \nabla - 2 i e {\cal A}))^{-1}
\Delta$
has to be evaluated.
 To do this we exploit the oscillatory character of the Abrikosov solution and
introduce 
raising and lowering operators
  $a=(\Lambda /\sqrt2)[\nabla_x + i ( \nabla_y -2 i e H x)]$ and 
$a^{\dagger} = 
               -(\Lambda / \sqrt2)[\nabla_x - i ( \nabla_y -2 i e H x)]$
obeying the usual bosonic commutation relations $[a,a^{\dagger}]=1$. 
If the
ground state equation (\ref{VL}) is denoted by $|0\rangle$, the higher eigenstates
(modes of the order parameter)
 are generated by
the standard procedure $a^{\dagger}|n\rangle =\sqrt{n+1}|n+1\rangle$.  
Then to make use of the properties of these operators the
operator ${\bf v}_f ( \nabla - 2 i e {\cal A})$
can  be rewritten as
$(v \sin \theta / \sqrt2 \Lambda) [ a e^{-i \phi} - a^{\dagger} e^{i \phi}]$
and the result of its acting on any state $|n \rangle$ evaluated
explicitly \cite{us}.
Using this approach
we are able to determine the unperturbed functions 
\begin{equation}
\label{g}
g=- {i }{\rm sgn}(\omega_n ) \Bigl[ 1- i \sqrt{\pi} 
                           \bigl({2 \Lambda \Delta \over v \sin\theta}\bigr)^2
                           W^{\prime}
                    \bigl ({2 i \widetilde \omega_n \Lambda {\rm sgn}(\omega_n)
                                     \over
                                       v \sin \theta}\bigr)\Bigr]^{-1/2}
\end{equation}
where $\Delta$ is the spatial average of the order parameter and
$W(z)=e^{-z^2}{\rm erfc}(-iz)$, and
\begin{equation}
\label{f}
\fl
f=2 i g {\sqrt{\pi} \Lambda \over v \sin\theta}
           \sum_{m=0}^{\infty}     
{1 \over \sqrt{m!}}\biggl( {-i \over \sqrt 2} \biggr)^m
 \bigl({\rm sgn} (\omega_n)\bigr)^{m+1} e^{i m \phi} 
W^{(m)}\biggl({2 i \widetilde \omega_n \Lambda {\rm sgn}(\omega_n)
                                    \over   
                    v \sin \theta}\biggr)  | m \rangle
\end{equation}
\noindent
 The expression for $g$ reproduces correctly
the gapped BCS-like function  for quasiparticles traveling
parallel to the magnetic field, while describing 
 gapless behaviour in all other directions.
A similar expression for $g$ has been obtained by
Pesch \cite{Pesch}.
Since $f$ is a Fourier series in $\phi$,
the mode with $m=0$ will 
couple to a scalar potential, the mode with $m=1$  to a transverse
potential etc. 
Then we find $\Delta_1=(i e A \Lambda \sqrt2) \bigl[
(1- i \bar \omega \tau)^{-1} + \omega_c \tau \bar\omega\tau \bigr]
| 1 \rangle$  (here $\bar \omega$ is the real external frequency). 
Using this value to determine the correction to anomalous propagator
$f_1$
and calculate the current from equation(\ref{curr}) we obtain the 
longitudinal and transverse conductivities up to order
$(\Lambda \Delta / v)^2$.

If we use the usual notation $\sigma_n$ for the normal
state conductivity $\sigma_n=N(0)e^2v^2\tau /3$,
the enhancement 
of the transverse current due to Lorentz force driven fluctuations 
of the order parameter
\begin{equation}
\sigma_{xy}^{fl}=6 \sigma_n \omega_c \tau (\Lambda \Delta / v)^2
\end{equation}
is exactly compensated by  the modification of the quasiparticle Hall current
due to additional scattering off the vortex lattice
\begin{equation}
\Delta\sigma_{xy}^{qp}=-6\sigma_n \omega_c \tau (\Lambda \Delta / v)^2.
\end{equation}
Similarly, the positive contribution to the
transverse conductivity due to the forces generated by 
gradient of the excited mode of 
the order
parameter 
\begin{equation}
\sigma_{xy}^{gr}=3\sigma_n (\Delta^2 \tau / E_f)
=12 \sigma_n \omega_c \tau (\Lambda \Delta / v)^2
\end{equation}
is cancelled by the additional scattering introduced by the deformed and
moving vortex lattice
\begin{equation}
\sigma_{xy}^{Th}=-12 \sigma_n \omega_c \tau (\Lambda \Delta / v)^2.
\end{equation}
As a result, the behaviour of the transverse conductivity $\sigma_{xy}$ is
determined solely by the effect of the modification 
of  the effective elastic scattering time $\tau_{eff}$
on the leading order
quasiparticle contribution.
For the dc conductivity this change is due to the decrease in the number of
states at the Fermi surface available for scattering as the 
superconducting gap opens. We
find, in agreement with the result of Pesch \cite{Pesch}
for the density of states, that the increase in the relaxation time
 is a non-analytic function
of the small parameter $(\Lambda \Delta / v)$
\begin{equation}
\label{tau}
\tau_{eff}^{-1}=\tau^{-1}\biggl[1 + 4 \biggl({\Lambda \Delta \over v}\biggr)^2
     	\log \biggl({\Lambda \Delta \over \sqrt 2 v}\biggr)
	+ 2 \biggl({\Lambda \Delta \over v}\biggr)^2 \biggr],
\end{equation}
and, up to
order $(\Lambda \Delta / v)^2$ the transverse conductivity is given by
\begin{equation}
\label{sxy}
\fl
\sigma_{xy}={1\over 3}N(0)e^2v^2 \tau_{eff} (\omega_c \tau_{eff})=
\sigma_n \omega_c \tau \biggl[ 1+ 4
                \biggl({\Lambda \Delta \over v}\biggr)^2
               \biggl(\log\biggl({2 v^2 \over\Lambda^2\Delta^2}\biggr)-1
			\biggr) \biggr].
\end{equation}
The longitudinal conductivity is obtained in a similar way
\begin{equation}
\label{sxx}
\sigma_{xx}=\sigma_n \biggl[ 1+2
                \biggl({\Lambda \Delta \over v}\biggr)^2
               \biggl(\log\biggl({2 v^2 \over\Lambda^2\Delta^2}\biggr)-1
			\biggr) \biggr].
\end{equation}
In the high-field region the 
square of the order parameter is linear in the 
applied magnetic field and is given by \cite{makit}
\begin{equation}
\label{d2}
\Delta^2 = { 1 \over  \pi N(0)} {H_{c2} - H \over \beta_A (2 \kappa_2^2 -1)}
           \biggl( H_{c2} - {T \over 2} {d H_{c2} \over dT} \biggr)
\end{equation}

In figure 1 we show the qualitative behaviour of $\sigma_{xy}$ as a function
of magnetic field, plotted using parameter values for Nb. 
The transverse conductivity is
enhanced below the upper critical field and has
negative curvature in the high field region.  
While the transverse conductivity is proportional to the 
square of the scattering 
time, the  Hall angle $\tan \theta_H=\sigma_{xy}/\sigma_{xx}$ 
is only linearly dependent on the scattering time and the corresponding
nonlinear dependence on magnetic field is weaker, as can be seen in
figure 2.
Finally, as the transverse resistivity $\rho_{xy}\approx \sigma_{xy}
/ \sigma_{xx}^2$ is independent of the effective scattering time, 
upon entering the superconducting state
it  remains 
linear in magnetic field with the same slope as in the normal metal.  
This behaviour is to be contrasted with
that of Bardeen-Stephen model \cite{BS}, where the resistivity is modified
but the Hall angle obeys the same linear law as in the normal state. 
On the other hand, the Nozieres-Vinen theory\cite{NV}, 
which predicts that the Hall angle should be constant in the flux-flow regime
below $H_{c2}$ at variance with the result of this work,
also finds that the transverse resistivity is identical to that
of the normal state, although the individual components
of the conductivity tensor are quite  different from those found here.

A comparison can be made with the experimental data of Fiory and Serin
\cite{Fiory} on high purity Nb. These experiments find
a transverse resistivity in the 
flux-flow regime which is linear in the applied magnetic field 
over a wide range of fields below $H_{c2}$.
The Hall angle, however, flattens or even increases above its value at
$H_{c2}$ before decreasing at lower fields.
 These results are more 
suggestive of the behaviour given here than the original interpretation 
given in terms of the Nozieres-Vinen theory. 
Also, the longitudinal resistivity found in \cite{Fiory}
has a distinct increase in slope just below the upper
critical field, which is consistent with the logarithmic behaviour 
given by equation (\ref{sxx}). 
Such comparisons are, of course, 
only 
qualitative, 
and more experimental work is needed to make a detailed comparison with the 
theory.

To conclude, we give the results of  a microscopic
calculation of the Hall resistivity 
of a clean type-II s-state
superconductor in the high-field 
limit. We find that the field dependence of the Hall conductivity 
in the high field regime, which is non-analytic, is
entirely due to the change in the density of quasiparticle states
at the Fermi level in the superconducting state. At the same time we
find that the field dependence  of the transverse resistivity 
below the upper critical field remains
unchanged. These results are in 
qualitative agreement with the experimentally observed behaviour.
The approach developed here can be generalized to 
superconductors with unconventional  order
parameter symmetry, this work is now in progress.

 One of us (A H) would like to thank A T Dorsey,
D Rainer, and K Scharnberg for important discussions.
This research was supported in part by the National Science Foundation
through Grant No DMR9008239.


\Figures
\begin{figure}
\caption{ Hall conductivity as a function of the reduced magnetic field}
\label{fig1}
\end{figure}

\begin{figure}
\caption{Hall angle as a function of the reduced magnetic field} 
\label{fig2}
\end{figure}
\stop
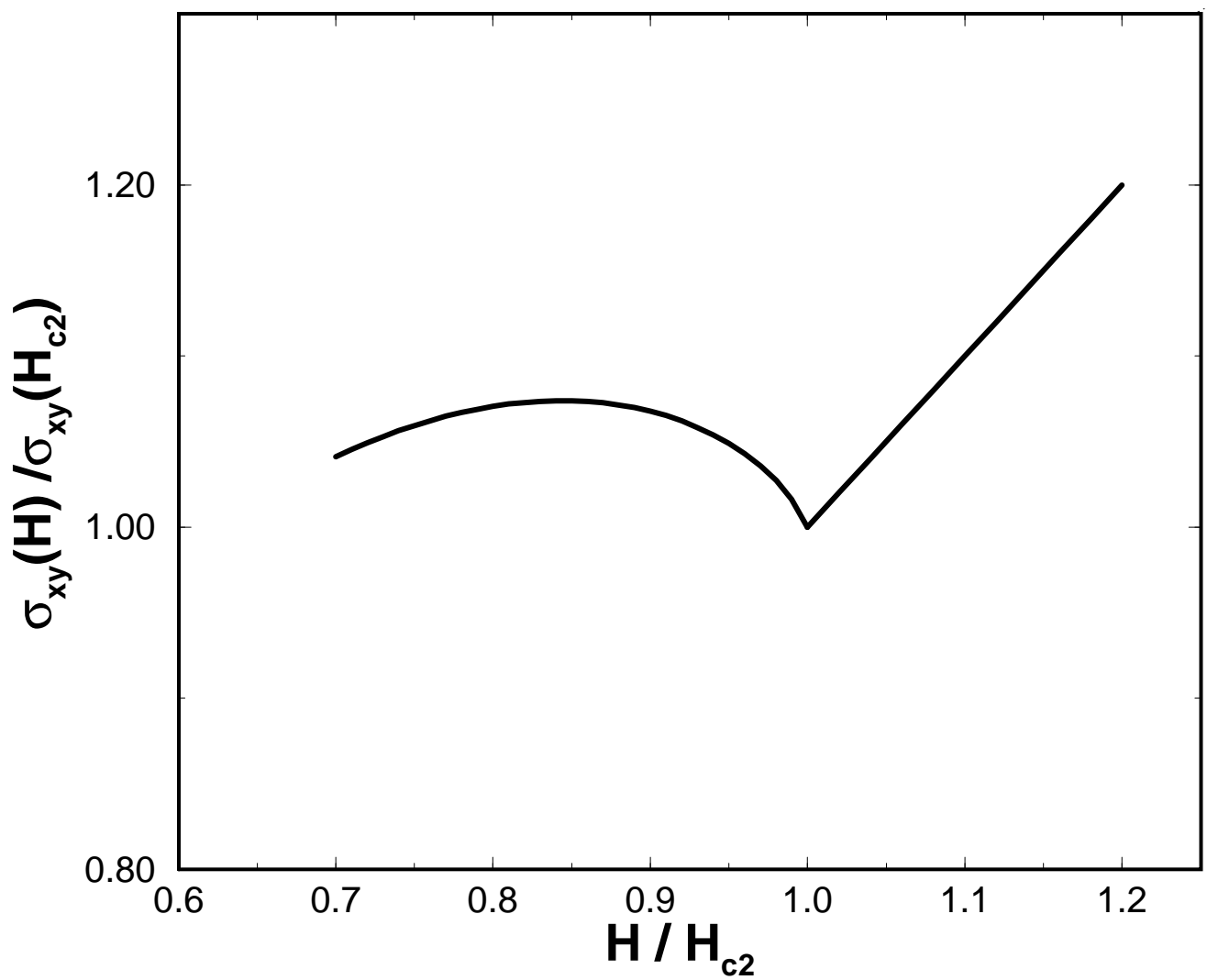
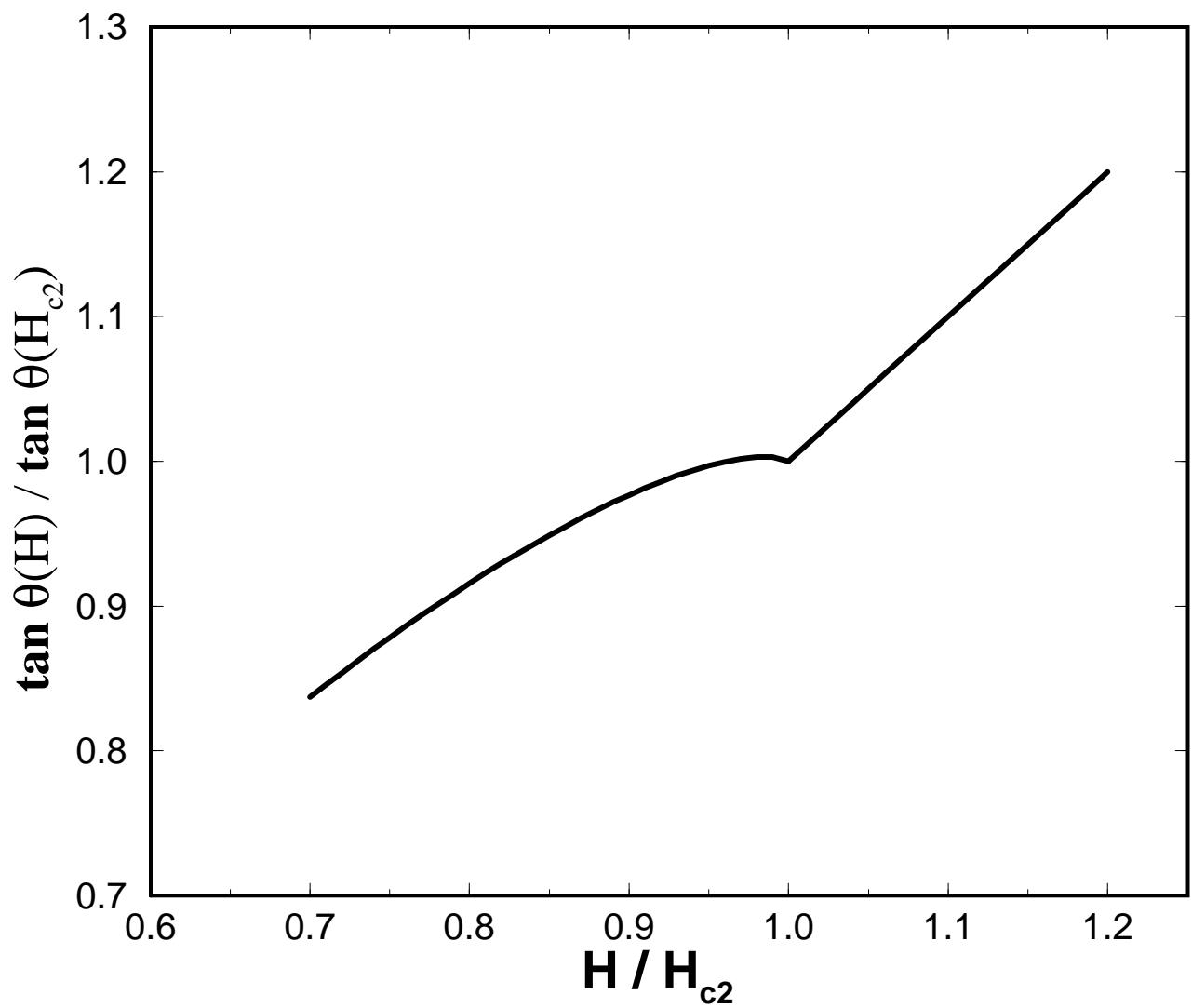
\end{document}